\documentclass[english,prl,aps,twocolumn,10pt]{revtex4-2}

\usepackage{amsmath}
\usepackage{amssymb}
\usepackage{graphicx}
\usepackage{babel}
\usepackage{array}
\usepackage{verbatim}
\usepackage[colorlinks=true, pdfstartview=FitV, linkcolor=blue, citecolor=blue, urlcolor=blue]{hyperref} 

\def\ie       {{\it i.e.}}
\def\cf       {{\it cf.}}
\def\via	  {{\it via \/}}

\newcommand{\mr}[1]{\mathrm{#1}}
\newcommand{\unit}[1]{\,\mathrm{#1}}

\newcommand{\um}{\mu m}

\newcommand{\phione}{\bar{\phi}_1}
\newcommand{\phitwo}{\bar{\phi}_2}

\newcommand{\phiuni}{\phi_{\mr{uc},i}}
\newcommand{\phiunone}{\phi_{\mr{uc},1}}
\newcommand{\phiuntwo}{\phi_{\mr{uc},2}}
\newcommand{\phicn}{\phi_\mr{c}}

\newcommand{\CPhis}{C_{\Phi_{1}\Phi_{2}}}
\newcommand{\Var}{V}
\newcommand{\COV}{\mr{Cov}}
\newcommand{\E}{E}

\newcommand{\ynv}{\gamma}
\newcommand{\thetaNVone}{\vartheta_{1}}
\newcommand{\phiNVone}{\varphi_{1}}
\newcommand{\thetaNVtwo}{\vartheta_{2}}
\newcommand{\phiNVtwo}{\varphi_{2}}

\newcommand{\Bc}{\vec{B}_\mr{c}}
\newcommand{\Bui}{\vec{B}_{\mr{uc},i}}

\newcommand{\vecB}{\vec{B}}
\newcommand{\vecBc}{\vec{B}_\mr{c}}
\newcommand{\vecBua}{\vec{B}_{\mr{uc},1}}
\newcommand{\vecBub}{\vec{B}_{\mr{uc},2}}
\newcommand{\vecBui}{\vec{B}_{\mr{uc},i}}
\newcommand{\vece}{\vec{e}}
\newcommand{\vecr}{\vec{r}}

\begin{document}
\title{Multiplexed scanning microscopy with dual-qubit spin sensors}	

\author{William S. Huxter$^{1,2}$, Federico Dalmagioni$^1$, and Christian L. Degen$^{1,2}$}
\affiliation{$^1$Department of Physics, ETH Zurich, Otto Stern Weg 1, 8093 Zurich, Switzerland.}
\affiliation{$^2$Quantum Center, ETH Zurich, 8093 Zurich, Switzerland}
\email{degenc@ethz.ch}

\begin{abstract}
	Scanning probe microscopy with multi-qubit sensors offers the potential to improve imaging speed and measure previously inaccessible quantities, such as two-point correlations. We develop a multiplexed quantum sensing approach with scanning probes containing two nitrogen-vacancy (NV) centers at the tip apex. A shared optical channel is used for simultaneous qubit initialization and readout, while phase- and frequency-dependent microwave spin manipulations are leveraged for de-multiplexing the optical readout signal.  Scanning dual-NV magnetometry is first demonstrated by simultaneously imaging multiple field projections of a ferrimagnetic racetrack device.
	Then, we record the two-point covariance of spatially correlated field fluctuations across a current-carrying wire.
	Our multiplex framework establishes a method to investigate a variety of spatio-temporal correlations, including phase transitions and electronic noise, with nanoscale resolution.
\end{abstract}
	
\date{\today}
\maketitle


The successes of quantum sensing with spin defects, particularly with nitrogen-vacancy (NV) centers in diamond, are supported by the high sensitivity, nanoscale resolution, and robust room-temperature functionality characteristic of these point-like sensors ~\cite{schirhagl14,janitz22}. When spin defects are integrated into scanning probe geometries~\cite{degen08apl,balasubramanian08}, simultaneous topography and stray-field imaging becomes possible. Such experiments have led to substantial advances when investigating nanoscale magnetic textures~\cite{finco23}, current distributions~\cite{ku20,jenkins22,palm24}, superconductivity~\cite{thiel16}, and even ferroelectricity~\cite{huxter23}. 

Thus far, the vast majority of scanning probe studies have relied on single-NV probes, owing to straightforward probe characterization and interpretation of measurement data.  The relatively few investigations using multi-NV scanning probes~\cite{laraoui15,tetienne16,simon21,liu23} have yet to fully explore the potential benefits in operating multiple spin sensors at the same time. For example, spin ensembles confined in nanoscale volumes present the opportunity to improve measurement speed or sensitivity while achieving spatial resolution below the optical diffraction limit. Intriguing prospects also lie in imaging quantities inaccessible to a single sensor, such as spatial correlations from field noise and fluctuations~\cite{rovny22,ji24}.
Two-point correlation measurements in other scanning probe microscopies~\cite{voigtlander18,kolmer19,leeuwenhoek20} require multiple tips and complex experimental setups, however, with spin defects in diamond, a single probe can easily host two or more individual sensors.  Moreover, methods are available to deterministically place defects within tens of nanometers distance~\cite{scarabelli16}.

\begin{figure*}
\includegraphics{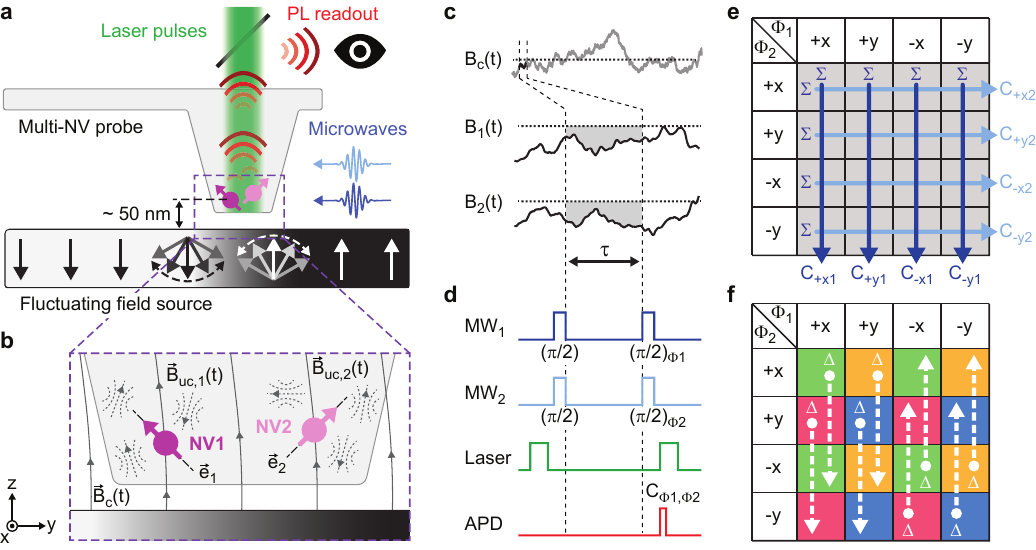}
\caption{\textbf{Concept of dual-NV scanning microscopy.}
	(a) Schematic of the scanning probe microscope. As the sample is moved under the probe, NV centers in the diamond tip sense their local magnetic (or electric) environment.
	A single optical channel is used for laser excitation and PL readout, while separate microwave frequencies simultaneously control both spins.
	(b) Detail showing the correlated ($\vecBc$, solid field lines) and uncorrelated ($\vecBua$, $\vecBub$, dashed) magnetic fields experienced by NV1 and NV2. The dashed black lines indicate the measurement axes $\vece_i$.
	(c) Time trace of the global magnetic signal $\vecBc(t)$ (arbitrary projection shown) and its projection onto every NV center ($B_i$).
	(d) Pulse sequence schematic. Laser pulses are used to polarize NV centers and prompt PL emission. Fixed microwave frequencies (MW$_i$) are used to simultaneously address a spin transition for each NV center. $\pi/2$ denotes the spin rotation angle and $\Phi_i \in \{0, 90^\circ, 180^\circ, 270^\circ\}$ denotes the relative phase shifts between first and second pulse.
	An avalanche photodiode (APD) converts the PL emission into sixteen photon counts signals $\CPhis$ that are used to recover the mean phases and count covariances in (e) and (f), respectively. 
	(e) Readout matrix for de-multiplexing of mean phases $\phione$ and $\phitwo$. Each matrix element represents a photon count $\CPhis$.  Mean phases are obtained by partial summation (arrows) followed by application of Eq.~(\ref{eq:mean_phases}).
	(f) Readout matrix for covariance sensing. Each matrix element represents the difference $V_{\Phi_{1}\Phi_{2}} - E_{\Phi_{1}\Phi_{2}}$, where $V$ and $E$ are the variance and expectation values computed across $n$ measurement repetitions, respectively.  Covariances are obtained by subtracting the readout combination connected by dashed arrows.  Four covariances can be computed, $\COV_{xx}$ (green), $\COV_{xy}$ (red), $\COV_{yx}$ (yellow), $\COV_{yy}$ (blue).}
\label{fig1}
\end{figure*}

In this work, we develop and realize scanning magnetometry experiments with diamond probes containing two active NV centers.  Our concept uses a single optical channel for spin state initialization and readout that is shared between NV centers, while phase-cycled microwave pulses allow for individual addressing of the two spins and de-multiplexing of the optical output signal into individual qubit responses.  Using a ferrimagnetic racetrack and a current-carrying wire as test devices, we demonstrate multiplexed imaging of magnetic field maps and covariance detection of spatially correlated field fluctuations with $\sim 50\unit{nm}$ spatial resolution.  The scanning probe geometry provides a unique test bed for the imaging of spatio-temporal correlations in general two-dimensional samples.

To introduce our sensing concept, we consider the general situation of two qubits exposed to the same time-varying magnetic field $\vecB(\vecr,t)$, see Fig.~\ref{fig1}.  Seen from the positions $\vecr_i$ of qubits $i=1,2$, the vector field presents two contributions: a global contribution $\Bc(t)$ that is experienced by both qubits, and local contributions $\Bui(t)$ that are different for each qubit (Fig.~\ref{fig1}(b)).
The field components $B_i(t)$ observed by the qubits are then given by
\begin{align}
	B_i(t) &= \vece_i \cdot \left( \vecBc(t) + \vecBui(t) \right),
\label{eq:bfield}
\end{align}
where $\vece_i$ (\cf~Fig.~\ref{fig1}(b)) is the projection axis of qubit $i$.
    
In our sensing experiment, the magnetic field is probed \via the accumulated phase difference in a time-evolved superposition of states~\cite{degen17}.  For a slowly varying magnetic field $B(t)$, the qubit-acquired phase is given by
\begin{align}
   	\phi_i(t) = \ynv \tau B_i(t),
\end{align}
where $\tau$ is the phase accumulation time and $\ynv=2\pi\times 28\unit{GHz/T}$ the NV gyromagnetic ratio~\cite{schirhagl14}.  To better understand how the correlated and uncorrelated magnetic fields impact multi-sensor measurements, we express the readout phases of two NV centers as
\begin{equation}
\begin{split}
	\phi_1(t) &= \phione + \phicn(t) + \phiunone(t) , \\
	\phi_2(t) &= \phitwo + m \phicn(t) + \phiuntwo(t) ,
\end{split}
\label{eq:phases}
\end{equation}
where $m = (\vece_2\cdot\vecBc) / (\vece_1\cdot\vecBc$).  Derived in the Supporting Information~\cite{supplemental}, Eq.~(\ref{eq:phases}) describes the relative phase accumulations for two separate superposition of states, where the total phases $\phi_i$ are composed of mean phases ($\bar{\phi}_i$), a correlated noise ($\phicn$ and $m\phicn$), and uncorrelated noises ($\phiuni$).  Together, $m$ and $\phicn$ characterize the strength of the fluctuations and the spatial correlation, which depends on the orientation and separations between the sensors.

Because the time-dependent terms average out across multiple measurements, only the mean phases $\bar{\phi}_i$ are usually measured in phased-based imaging experiments~\cite{ku20,huxter23,palm24}. However, the noise term $\phicn(t)$ generates a non-zero covariance between sensors accessible across many repeated measurements.  This covariance signal contains important information about the spatial-temporal correlations of the magnetic field~\cite{rovny22}.
A scheme able to measure both the mean phases and signal correlations across multiple sensors would therefore be a valuable asset in the metrology of spatially correlated signals.
	
Our implementation for such a scheme is presented in Fig.~\ref{fig1}(d).  The pulse sequence uses two fixed frequencies -- each tuned to the spin resonance of one NV center -- that are applied at the same time~\cite{schloss18}.  Sixteen unique microwave phase readout combinations (denoted by $\Phi_1$ and $\Phi_2$) then permit separating both NV center contributions in the summed photoluminescence (PL) signal
\footnote{In principle, only eight phase combinations are needed to separate NV center contributions, see SM~\cite{supplemental}.}.
This readout scheme is capable of measuring both static fields (with the Ramsey protocol shown in Fig.~\ref{fig1}(d)), coherent AC fields (with dynamical decoupling protocols~\cite{gullion90,palm22,huxter22}), and spatial correlations through covariance measurements (details below).  Moreover, the scheme can be readily generalized to more than two sensors~\cite{supplemental}.  The only requirement is that spin transitions are spectrally separated to minimize spurious excitation or dynamical frequency shifts~\cite{bloch40}. When spin transitions for the sensors are degenerate, a subset of the sixteen-phase readout scheme can be used to sense the mean phase difference $|\phione - \phitwo|$~\cite{supplemental}.

After $n$ measurement repetitions, the PL counts for a given phase combination have the form
\begin{equation}
\begin{split}
	\CPhis &= nc_{1} \left(1 - \frac{\epsilon_1}{2}\left[1  +  e^{-\zeta_1(\tau)}\cos(\phione + \Phi_1)  \right] \right) \\
	&+nc_{2} \left(1 - \frac{\epsilon_2}{2}\left[1  +  e^{-\zeta_2(\tau)}\cos(\phitwo + \Phi_2)  \right] \right),
\end{split}
\label{eq:counts}
\end{equation}
where $c_i$, $\epsilon_i$, $e^{-\zeta_i(\tau)}$ denote the average number of photons detected per readout for the $m_S=0$ spin state, the optical contrast, and the exponential spin dephasing during $\tau$ for each NV center, respectively~\cite{palm22,supplemental}.
Typical values for our experiments are $c_i \sim 0.1$, $\epsilon_i \sim 0.2$, $\tau\sim 250 \unit{ns}$, $\zeta_i(\tau) \sim 0.7$ and $n \sim 10^{5}-10^{8}$.
$\Phi_i$ represent the four qubit axes in the $xy$ plane of the Bloch sphere and are denoted as $\Phi \in \{+x, +y, -x, -y\}$~\cite{ku20,palm22}. 

The mean phases $\phione$ and $\phitwo$ can be extracted by combining partial sums of the PL counts, defined as $C_{\Phi_1} = \sum_{\Phi_2} \CPhis$ and $C_{\Phi_2} = \sum_{\Phi_1} \CPhis$, indicated by the vertical and horizontal blue arrows in the readout matrix of Fig.~\ref{fig1}(e).
For the Ramsey protocol, the phase computations are
\begin{equation}
	\phi_i =  \arctan\left(\frac{C_{-y_i} - C_{+y_i}}{C_{-x_i} - C_{+x_i}} \right). 
\label{eq:mean_phases}
\end{equation}
In Eqs.~(\ref{eq:counts}) and (\ref{eq:mean_phases}), each measurement repetition measures both NV centers and every phase combination is used to compute the accumulated mean phases. This establishes a multiplexed framework for simultaneously controlling and reading out multiple non-interacting qubits over a single optical channel.

To measure correlations, the Poissonian nature of the single-photon emission process during optical readout must be considered.  Since the expected number of photons per readout $\E$, and its variance $\Var$, are related to the phase accumulated, any covariance between NV centers (resulting from $\phicn$ and $m\phicn$, see Eq.~(\ref{eq:phases})), is found in the variance of the summed photon counts.  The four readout phase combinations most sensitive to phase accumulation, $\Phi_1\Phi_2 = \pm y \pm y$,  can be turned into a count covariance, $\COV_{yy} = \langle \sin\phi_1 ,\sin\phi_2 \rangle \sim \langle \phi_1, \phi_2 \rangle$, by combining the expectation values $\E_{\pm y \pm y}$ and variances $\Var_{\pm y \pm y}$ taken across many multiplexed readouts (see Fig.~\ref{fig1}(f))~\cite{supplemental}. Following the law of total covariance, one of the two equivalent expressions for $\COV_{yy}$ is
\begin{align}
	\COV_{yy}&=\frac{(\Var_{+y+y}-\E_{+y+y})-(\Var_{+y-y}-\E_{+y-y})}{c_{r1}c_{r2}} .
\label{eq:cov}
\end{align}
Here, $c_{ri} = c_i \epsilon_i e^{-\zeta_i(\tau)}$ is the readout contrast of each NV center~\cite{supplemental}.  Three other count covariances ($\COV_{xx}$, $\COV_{xy}$, and $\COV_{yx}$) can be computed from the remaining sixteen-phase readouts through analogous expressions to Eq.~(\ref{eq:cov}), see color coding in Fig.~\ref{fig1}(f)~\cite{supplemental}.
In general, all four count covariances are needed to isolate the covariance of the correlated phases $\phicn$ and $m\phicn$, since non-zero $\phione$ and $\phitwo$ mix the correlation among different count covariances~\cite{supplemental}. However, when $\phione \approx \phitwo \approx 0$, $\COV_{yy}$ contains the entire correlation signal: 
\begin{align}
	\COV_{yy} = \langle\sin(\phicn),\sin (m \phicn)\rangle \sim m \phicn^2 .
\end{align}
	
\begin{figure}[!tb]
\includegraphics[width=1.00\columnwidth]{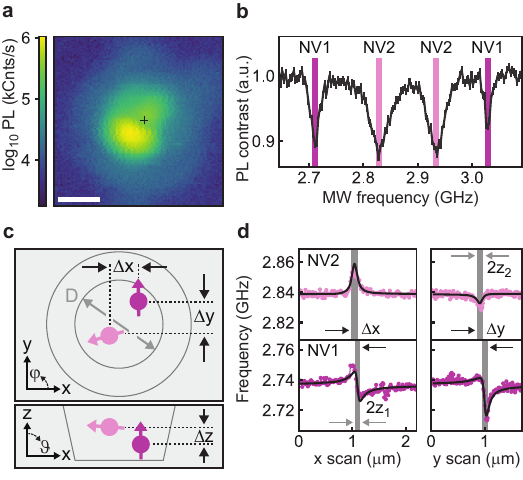}
\caption{\textbf{Characterization of dual-NV probes.}
	(a) Log-scale confocal image of the collected PL recorded by scanning the excitation laser across the diamond probe. Multiple bright spots indicate multiple NV centers and a fainter halo outlines the pillar structure.  Scale bar, $1\unit{\um}$.
	(b) Optically-detected magnetic resonance (ODMR) spectra identifying four spin transitions, confirming the presence of two NV centers. The spectra are recorded while focusing the laser at the position indicated by the cross in panel (b).
	(c) Schematic of two NV centers inside the scanning tip showing their orientation (defined by $\vartheta$ and $\varphi$) and relative positions (defined by $\Delta x$, $\Delta y$, and $\Delta z$).
	(d) ODMR frequency shifts while scanning across orthogonal step edges of a magnetic stripe. $\Delta x$ and $\Delta y$ (indicated with black arrows) are determined from the position of the fitted edges of the sample (gray lines). The gray line widths (indicated with gray arrows) is twice the fitted standoff distance ($z_1=47\pm1\unit{nm}$ and $z_2=58\pm2\unit{nm}$ for the plotted data), which determines $\Delta z$ and the spatial resolution.
	Data recorded on tip \#2.}
\label{fig2}
\end{figure}
\begin{figure*}[!tb]
\includegraphics{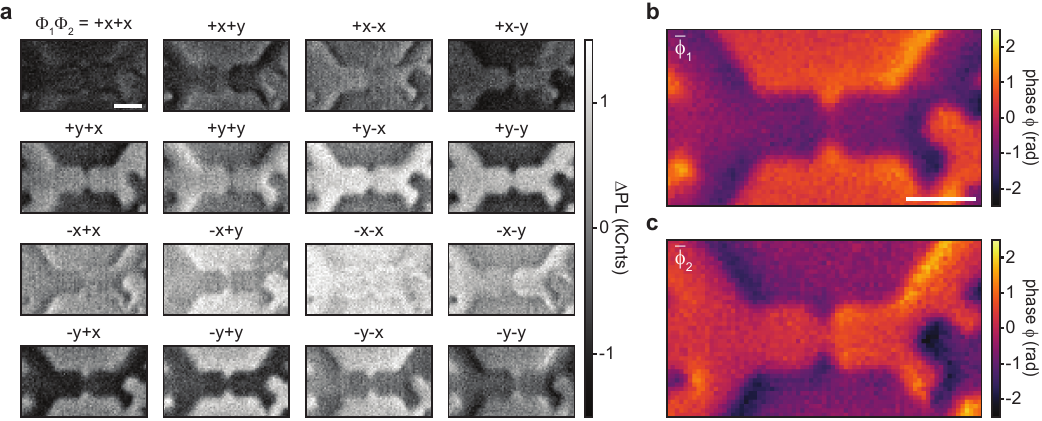}
\caption{\textbf{Multiplexed imaging of a ferrimagnetic GdCo racetrack.}
	(a) Experimental PL count data, $\CPhis$, for the sixteen readout combinations forming the readout matrix in Fig.~\ref{fig1}(e). The mean PL across all phase combinations has been subtracted.
	(b,c) Phase images $\phione$ (NV1) and $\phitwo$ (NV2), computed using Eq.~(\ref{eq:mean_phases}).  
	Scale bars, $2\unit{\um}$.
	Data recorded on tip \#4.}
\label{fig3}
\end{figure*}

We experimentally demonstrate multiplexed and correlated sensing using the arrangement sketched in Fig.~\ref{fig1}(a).  Following typical scanning NV platforms~\cite{degen08apl,balasubramanian08}, a sample stage is scanned underneath a diamond probe where laser pulses, microwave pulses, and single-photon counting are used for spin control and readout.  Extending single-NV scanning microscopy to dual-NV (or multi-NV) microscopy therefore requires no change in instrumentation. Dual-NV probes (QZabre AG) are fabricated by adjusting dopant concentrations during nitrogen implantation~\cite{orwa11,oforiokai12} and post-selecting those containing two NV centers. 

With two NV centers in the same nanophotonic structure (\ie, the diamond tip), laser irradiation interacts with both NV centers, coupling their optical excitation and photon collection.  When a laser is scanned across the diamond probe, multiple bright regions are observed (Fig.~\ref{fig2}(a)), reflecting the multimodal nature of the tips used in this work~\cite{zhu23}. By contrast, a single-mode structure would only produce a single PL maximum regardless of the number of emitters. 
Due to the complex spatial structure of the laser mode volume, pillar shape and random NV center location, attributing a bright spot in Fig.~\ref{fig2}(a) to a certain NV center is not possible.  Instead, each laser position excites one or both NV centers with varying efficiencies ~\cite{supplemental}.  Positioning the laser focus to illuminate both NV centers (black cross in Fig.~\ref{fig2}(a)) and sweeping the microwave excitation frequency reveals four PL dips in Fig.~\ref{fig2}(b). These dips correspond to the two allowed spin transitions for each NV center~\cite{schirhagl14,janitz22,doherty13}. Since the NV centers in this probe have different orientations ($\vece_1 \nparallel \vece_2$), a small magnetic bias field spectrally separates each transition.
	
In addition to having different orientations, NV centers also have different spatial locations $\vecr_1$ and $\vecr_2$ as shown schematically in Fig.~\ref{fig2}(c).  The spatial separation is between $0-D$ in the $xy$ plane, where $D\sim 300-400\unit{nm}$ is the tip apex diameter~\cite{zhu23}, and $\sim 5-20\unit{nm}$ in the $z$ direction, owing to the $\sim 5\unit{keV}$ ion implantation energy~\cite{oforiokai12}.  To properly interpret measurement data the orientations and relative distance vector $\vecr_2-\vecr_1 = (\Delta x, \Delta y, \Delta z)$ must be determined experimentally.
To do so, we scan the tip across perpendicular edges of a magnetic calibration sample~\cite{hingant15} and fit the ODMR profiles by a magnetic stray field model~\cite{supplemental}.  This procedure gives estimates for the orientations of both NV centers, their lateral separation, and their respective $z$ standoff distances.
Fig.~\ref{fig2}(d) plots example line scan data with fits for two of the four measured spin transitions.  For the probe characterized in Fig.~\ref{fig2}, the NV orientations are $\vece_1=(\thetaNVone,\phiNVone) = (41\pm2^\circ, 94\pm2^\circ)$ and $\vece_2=(\thetaNVtwo,\phiNVtwo) = (-84\pm2^\circ, -161\pm2^\circ)$, and their separation is $\Delta x = -52\pm7\unit{nm}$, $\Delta y = -96\pm6\unit{nm}$, $\Delta z = 11\pm5\unit{nm}$ in the laboratory frame.  Characterization data for three further probes are given in~\cite{supplemental}.
	 
We demonstrate multiplexed imaging on two representative magnetic samples.  The first sample is a ferrimagnetic GdCo racetrack device~\cite{liu23prb} that is uniformly magnetized in the out-of-plane direction~\cite{supplemental}.
Imaging measurements for the sixteen-phase readout scheme are shown in Fig.~\ref{fig3}(a).  De-multiplexing of the PL count images (Fig.~\ref{fig3}(b,c)) reproduces the individual phase maps $\phione$ and $\phitwo$, representing different projections of the stray field vector onto the individual axes $\vece_1$ and $\vece_2$ of NV1 and NV2, respectively. In addition, the images are slightly shifted because the NV centers sit at different locations in the tip ($\sqrt{\Delta x^2+\Delta y^2}\sim 180\unit{nm}$ for this tip).
The ability to simultaneously provide multiple vector field projections is one advantage of multi-qubit probes~\cite{broadway20}.  Opposite to ensemble probes~\cite{liu23}, however, our individual NV readout maintains the nanoscale spatial resolution offered by single spin probes.  Additionally, experiments show that simultaneous operation of both NV centers improves measurement sensitivity by approximately $\sqrt{2}$, as expected from the $\propto\sqrt{N}$ scaling for $N$ sensors~\cite{degen17,supplemental}.

\begin{figure}[!tb]
\includegraphics{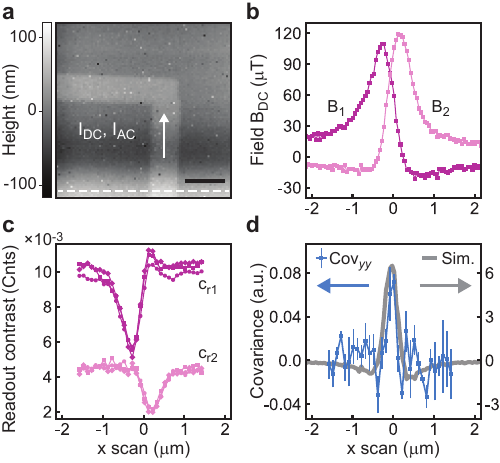}
	\caption{\textbf{Sensing of spatial correlations using covariance detection.}
	(a) Topography of the current-carrying wire device. Scale bar, $1\unit{\um}$.
	(b) Dual-NV magnetometry line scans taken across the wire (white-dashed line in (a)). A DC current through the wire produces a static magnetic field.
	(c) Phase contrasts $c_{r1}$ and $c_{r2}$ for three covariance imaging measurements where an asynchronous AC current ($f = 35.211430\unit{kHz}$) is passed through the wire.  The dips in the signal reflect the magnetic field noise seen by each NV center leading to decoherence. Data point shapes (square, circle, and diamond) indicate different experimental runs. 
	(d) Covariance signal computed from the averaged measurements of (c). Error bars are the standard deviation between the measurements.  Gray curve is a Monte Carlo simulation of the expected phase covariance $\COV_{yy}$ using the measurements in panel (b) as input.
	Data recorded on tip \#4.}
\label{fig4}
\end{figure}

For our second experiment we use a current-carrying wire~\cite{chang17} consisting of a $\sim 700\unit{nm}$ wide Ti-Au stripe, see Fig.~\ref{fig4}(a)~\cite{supplemental}.
Static and fluctuating magnetic fields are generated by injecting DC or asynchronous AC current waveforms.  Figure~\ref{fig4}(b) displays the field profiles for DC injection observed in a line scan taken across the width of the wire.  Similar to Fig.~\ref{fig3}(b,c), multiplexing is able to record separate profiles for NV1 and NV2 reflecting the differences in spin orientations and positions.

To demonstrate covariance imaging, we inject an AC current that runs asynchronous to the measurements sequence. By choosing a frequency $f$ much slower than the duration of a single phase measurement ($f \sim 35\,\mr{kHz} \ll \tau^{-1} = 4\,\mr{MHz}$), individual phase readouts can be treated like random samples from an arcsine distribution.
A correlated phase component $\phicn(t)$ [Eq.~(\ref{eq:phases})] is naturally formed since the fluctuating current in the wire produces the fields observed by both NV centers.
	
Figure~\ref{fig4}(c,d) displays the PL count data from three asynchronous AC scanning experiments. The phase contrasts $c_{r1}$ and $c_{r2}$ in Fig.~\ref{fig4}(c) show dips that indicate the root-mean-square (rms) magnetic noise observed by each NV center. Similar decoherence-based measurements are used in the detection of fluctuating spins~\cite{cole09,loretz14apl}. The shapes of the dips mimic the static field measurements in Fig.~\ref{fig4}(b), since the rms noise of an oscillating signal is proportional to the amplitude of oscillation. Different background levels for $c_{r1}$ and $c_{r2}$ are due to changes in the laser focus position.
The averaged count covariance, $\COV_{yy}$, is shown in Fig.~\ref{fig4}(d). The singly-peaked shape of the data, which corresponds to the region where both NV centers see strong fluctuations with the same polarity, agree well with the covariance profile estimated from the DC data (gray curve)~\cite{supplemental}. With negligibly small static fields in the measurements, the entire correlated signal is found in the $\COV_{yy}$ measurement.
	
The relatively low signal-to-noise ratio (SNR) in the covariance data (Fig.~\ref{fig4}(d)) is due to a combination of low initialization and readout fidelities as well as the presence of uncorrelated noise causing additional decoherence, leading to $c_{ri}\sim 10^{-3} \ll 1$ (Fig.~\ref{fig4}(c)).  Since the SNR depends quadratically on $c_{ri}$~\cite{rovny22}, use of improved laser initialization and readout protocols~\cite{song20,wirtitsch23}, photonic wave guiding~\cite{aharonovich14,zhu23} and charge-state or logic-based readouts~\cite{hopper18,rovny22} could strongly boost sensitivity.
Substantial SNR gains could also be made by extending the available free evolution time $\tau$ through dynamical decoupling~\cite{ryan10} since the covariance signal is proportional to $\phicn^2 \propto \tau^2$.  Dynamical decoupling could further be used to decipher the frequency spectra of noise sources~\cite{alvarez11}.  Alternatively, relative time shifts between pulses could be introduced to analyze temporal correlations down to nanosecond time scales~\cite{herb24expt}, offering perspectives for probing fast spatio-temporal correlations.

Looking forward, the ability to measure multiple qubits simultaneously, including the spatial correlation among qubits, creates the possibility to investigate a variety of correlated phenomena with multi-NV scanning probes. This includes spatial phase transitions in magnetic, superconducting, and ferroelectric materials~\cite{vaterlaus00,hilgenkamp02,li21}, as well as correlated electronic noise in conductors~\cite{agarwal17,zhang24}, including hydrodynamic flow~\cite{mendoza11,gabbana18}.
Additionally, while our experimental demonstration used two non-interacting qubits, our framework is scalable to more sensors and can serve as a starting point for imaging higher order correlations and fluctuations~\cite{ji24} or for sensing with entangled states.  The latter approach, where sensor separations $\lesssim 30\unit{nm}$ would be needed~\cite{neumann10natphys}, could enable sensing beyond the standard quantum limit~\cite{degen17,xie21}.


\begin{acknowledgments}
	The authors thank K. Herb, N. Meinhardt, N. Prumbaum, and S. Singh for helpful discussions, Z. Liu and A. Hrabec for providing the GdCo ferrimagnet samples, G. Puebla-Hellmann for providing the wire sample, C. Ding for wire bonding, and J. Rhensius for support in scanning probe fabrication. 
	This work was supported by the European Research Council under ERC CoG 817720 (IMAGINE), by the Swiss National Science Foundation (SNSF) under Grants No. 200020\_212051, 200021\_219386, IZRPZ0\_194970 and CRSII\_222812, and by the State Secretariat for Education, Research, and Innovation (SBFO) under Grant No. UeM019-8.
\end{acknowledgments}

\end{document}